\documentstyle[12pt]{article}

\font\twlgot =eufm10 scaled \magstep1
\font\egtgot =eufm8
\font\sevgot =eufm7

\font\twlmsb =msbm10 scaled \magstep1
\font\egtmsb =msbm8
\font\sevmsb =msbm7

\newfam\gotfam

\textfont\gotfam\twlgot
\scriptfont\gotfam\egtgot
\scriptscriptfont\gotfam\sevgot

\newfam\msbfam
\textfont\msbfam\twlmsb
\scriptfont\msbfam\egtmsb
\scriptscriptfont\msbfam\sevmsb

\def\pBbb{\relax\ifmmode\expandafter\Bb\else\typeout{You cann't use
Bbb in text mode}\fi}
\def\Bb #1{{\fam\msbfam\relax#1}}

\def\thebibliography#1{\bigskip\section*{\centering
References\\}\bigskip\list
  {\arabic{enumi}.}{\settowidth\labelwidth{#1}\leftmargin\labelwidth
    \advance\leftmargin\labelsep
    \usecounter{enumi}}
    \def\newblock{\hskip .11em plus .33em minus .07em}
    \sloppy\clubpenalty4000\widowpenalty4000
    \sfcode`\.=1000\relax}

\newcommand{\Si}{\Sigma}

\def\op#1{\mathop{\fam0 #1}\limits}

\newcommand{\Id}{{\rm Id\,}}
\def\Ker{{\rm Ker\,}}

\newcommand{\ben}{\begin{eqnarray}}
\newcommand{\een}{\end{eqnarray}}
\newcommand{\be}{\begin{eqnarray*}}
\newcommand{\ee}{\end{eqnarray*}}
\newcommand{\bea}{\begin{eqalph}}
\newcommand{\eea}{\end{eqalph}}

\newcommand{\cD}{{\cal D}}
\newcommand{\cL}{{\cal L}}

\newcommand{\cE}{{\cal E}}

\newcommand{\la}{\lambda}

\newcommand{\m}{\mu}

\newcommand{\g}{\gamma}
\newcommand{\G}{\Gamma}
\newcommand{\e}{\epsilon}

\newcommand{\si}{\sigma}

\newcommand{\w}{\wedge}

\newcommand{\wt}{\widetilde}
\newcommand{\wh}{\widehat}
\newcommand{\ol}{\overline}

\newcommand{\dr}{\partial}

\newcounter{eqalph}
\newcounter{equationa}

\newenvironment{eqalph}{\stepcounter{equation}
\setcounter{equationa}{\value{equation}}
\setcounter{equation}{0}

\begin{eqnarray}}{\end{eqnarray}
\setcounter{equation}{\value{equationa}}}

\begin{document}
\hbox{}

\centerline{\bf\large GRAVITY AS A HIGGS FIELD.}
\medskip

\centerline{\bf\large I. GEOMETRIC EQUIVALENCE PRINCIPLE}
\bigskip

\centerline{\bf Gennadi A Sardanashvily}
\medskip

\centerline{Department of Theoretical Physics, Physics Faculty,}

\centerline{Moscow State University, 117234 Moscow, Russia}

\centerline{E-mail: sard@theor.phys.msu.su}
\bigskip

\centerline{\bf Abstract}
\medskip

{\it If gravity is a metric field by Einstein, it is a Higgs field.}
Gravitation theory meets spontaneous symmetry breaking in accordance
with the equivalence principle reformulated in the spirit of Klein-Chern
geometries of invariants. In gravitation theory, the structure group of
the principal linear frame bundle $LX$ over a world manifold $X^4$ is
reducible to the connected Lorentz group $SO(3,1)$. The physical
underlying reason of this reduction is Dirac fermion matter possessing
only exact Lorentz symmetries. The associated Higgs field is a tetrad
gravitational field $h$ represented by a global section of the quotient
$\Si$ of $LX$ by $SO(3,1)$. The feature of gravity as a Higgs field
issues from the fact that, in the presence of different tetrad fields,
Dirac fermion fields are described by spinor bundles associated with
different reduced Lorentz subbundles of $LX$, and we have nonequivalent
representations of cotangent vectors to $X^4$ by Dirac's matrices. It
follows that a fermion field must be regarded only in a pair with a
certain tetrad field. These pairs fail to be represented by sections of
any product bundle $S\times\Si$, but sections of the composite spinor
bundle $S\to\Si\to X^4$. They constitute the so-called fermion-gravitation
complex where values of tetrad gravitational fields play the role of
coordinate parameters, besides the familiar world coordinates. In Part 1 of
the article, geometry of the fermion-gravitation complex is investigated.
The goal is the total Dirac operator into which components of a connection
on $S\to\Si$ along tetrad coordinate directions make contribution. The
Part II will be devoted to dynamics of fermion-gravitation complex. It is
a constraint system to describe which we use the covariant multisymplectic
generalization of the Hamiltonian formalism when canonical momenta
correspond to derivatives of fields with respect to all world coordinates,
not only the time.

\section{Introduction}

Gravitation theory is theory with spontaneous symmetry breaking.
Spontaneous symmetry breaking is quantum phenomenon modelled by a classical
Higgs field. In the algebraic quantum field theory, Higgs fields
characterize nonequivalent Gaussian states of algebras of quantum fields
\cite{nsar}. They are {\it sui generis} fictitious fields describing
collective phenomena.
In gravitation theory, spontaneous symmetry breaking displays on the
classical level. It is established by the equivalence principle reformulated
in the terms of Klein-Chern geometries of invariants \cite{iva,tsar,3sar}.

In Einstein's General Relativity, the equivalence principle is called to
provide transition to Special Relativity with respect to some reference
frames. In the spirit of F.Klein's Erlanger program, the Minkowski space
geometry can be characterized as geometry of Lorentz invariants. The
geometric equivalence principle then postulates that there exist reference
frames with respect to wich Lorentz invariants can be defined everywhere
on a world manifold $X^4$. This principle has
the adequate mathematical formulation in terms of fibre bundles.

We follow the generally accepted geometric description of classical fields
as sections of a fibred manifold
\[
\pi: Y\to X
\]
over a world manifold $X^4$. In gauge theory, $Y\to X$ is a bundle
with a structure group.

Let $LX$ be the principal bundle of linear frames in tangent spaces to
$X^4$. Its structure group is
\[
GL_4=GL^+(4,{\bf R}).
\]
The geometric equivalence principle requires that this structure group
is reduced to the connected Lorentz group
\[
L=SO(3,1).
\]
It means that there is given a reduced subbundle $L^hX$ of $LX$ whose
structure group is $L$. They are atlases of $L^hX$ with respect to which
Lorentz invariants can be defined.

In accordance with the well-known theorem, there is
the 1:1 correspondence between the reduced $L$ subbundles $L^hX$ of
$LX$ and the tetrad gravitational fields $h$ represented by global
sections of the quotient bundle
\begin{equation}
\Si=LX/L\to X^4. \label{5.15}
\end{equation}
Its standard fibre is the quotient space $GL_4/L$. The bundle
(\ref{5.15}) is isomorphic to the 2-fold covering of the bundle $\Si_g$
of pseudo-Riemannian forms in cotangent spaces to $X^4$. A global section
of $\Si_g$ is a pseudo-Riemannian metric on $X^4$.

Thereby, the geometric equivalence principle provides a world manifold
with the so-called $L$-structure \cite{sul}. From the physical point of
view, it singles out the Lorentz group as the exact symmetry subgroup
of world symmetries broken spontaneously \cite{iva}. The associated
classical Higgs field is a tetrad (or metric) gravitational field.

For the first time, the conception of a graviton as a Goldstone particle
corresponding to violation of Lorentz symmetries in a curved space-time
had been advanced in mid 60s by Heisenberg and Ivanenko in discussion on
cosmological and vacuum asymmetries. This idea was revived in connection
with constructing the induced representations of the group $GL_4$
\cite{ish,nee,ogi} and then in the framework of the approach to
gravitation theory as a nonlinear $\si$-model \cite{per}. In geometric
terms, the fact that a pseudo-Riemannian metric is similar a Higgs field
has been pointed out by Trautman \cite{tra} and by us \cite{0sar}. To
prove it, the new geometric formulation of the equivalence principle
has been suggested \cite{0iva,iva}.

The underlying physical reason of the geometric equivalence principle
is Dirac fermion matter possesing only exact Lorentz symmetries.

Let us consider a bundle of complex Clifford algebras ${\bf C}_{3,1}$
over $X^4$. Its subbundles are both a spinor bundle $S_M\to X^4$ and the
bundle $Y_M\to X^4$ of Minkowski spaces of generating elements of
${\bf C}_{3,1}$. There is the bundle morphism
\[
\g: Y_M\otimes S_M\to S_M
\]
which defines representation of elements of $Y_M$ by Dirac's
$\g$-matrices on elements of the spinor bundle $S_M$. To describe Dirac
fermion fields on a world manifold, one must require that the bundle
$Y_M$ is isomorphic to the cotangent bundle $T^*X$ of $X^4$. It takes
place if $Y_M$ is associated with some reduced $L$ subbundle $L^hX$
of the linear frame bundle $LX$. Then, there exists the representation
\[
\g_h : T^*X\otimes S_h\to S_h
\]
of cotangent vectors to a world manifold $X^4$ by Dirac,s $\g$-matrices
on elements of the spinor bundle $S_h$ associated with the lift of $L^hX$
to a $SL(2,{\bf C})$ principal bundle. Sections of $S_h$ describe
Dirac fermion fields in the presence of a tetrad gravitational field $h$.

It follows that, in the presence of Dirac fermion matter, we must handle
two types of reference frames. They are holonomic atlases of $LX$ and
atlases of a reduced Lorentz subbundle $L^hX$ of $LX$. The corresponding
tetrad garvitational field $h$ determines transformations between these
reference frames. The percularity of gravity thus is clarified. In contrast
to other fields, a tetrad gravitationsl field itself defines reference
frames. The Higgs field character of gravity issues from the fact that
the reference frames and other characteristics corresponding to different
gravitational fields are not equivalent in a sense.

The key point of our consideration consists in the fact that,
for different tetrad fields $h$ and $h'$, the representations $\g_h$ and
$\g_{h'}$ fail to be equivalent \cite {tsar,3sar}. It follows that every
Dirac fermion field must be regarded only in a pair with a certain
tetrad gravitational field $h$. These pairs constitute the so-called
fermion-gravitation complex \cite{nee}. They can not be represented by
sections of any product $S\times\Si$ where $S\to X^4$ is some standard
spinor bundle. At the same time, there is the 1:1 correspondence between
these pairs and the sections of the composite bundle
\begin{equation}
S\to\Si\to X^4 \label{L1}
\end{equation}
where $S\to\Si$ is a spinor bundle associated with the $L$ principal
bundle $LX\to\Si$ \cite{2sar,3sar}. In particular, every spinor bundle
$S_h\to X^4$ is isomorphic to restriction of $S$ to $h(X^4)\subset \Si$.

To show the physical relevance of the composite bundle (\ref{L1}), we
aim constructing the total Dirac operator $\cD$ on its section.

We however observe that the
composite bundle (\ref{L1}) fails to possess a structure group.
Therefore, connections
on $S\to X^4$ which we consider are not principal connections. They are
so-called composite connections constructed by means of principal
connections on $S\to\Si$ and $\Si\to X^4$ \cite{2sar,3sar,sard}. To
handle them, we follow the general notion of connections as sections
of jet bundles.

Recall that the $k$-order jet manifold $J^kY$ of a fibred
manifold $Y\to X$ comprises the equivalence classes
$j^k_xs$, $x\in X$, of sections $s$ of $Y$ identified by the $(k+1)$
terms of their Taylor series at $x$. It is a finite-dimensional manifold.
Jet manifolds have been widely used in the
theory of differential operators. Their application to differential
geometry is based on the  1:1 correspondence between the connections
on a fibred manifold $Y\to X$ and the global sections of the jet bundle
$J^1Y\to Y$ \cite{man,sard,sau}.

Dynamics of fields represented by sections of a fibred
manifold $Y\to X$ is phrased in terms of jet manifolds
\cite{gia,got,kup,sard}. In the first order Lagrangian formalism,
the jet manifold $J^1Y$ plays the role of a finite-dimensional
configuration space of fields. Given fibred coordinates $(x^\la,y^i)$
of $Y\to X$, it is endowed with the adapted
coordinates $(x^\la,y^i,y^i_\la)$ where coordinates $y^i_\la$ make the
sense of values of partial derivatives $\dr_\la y^i(x)$ of field
functions $y^i(x)$. A Lagrangian density on $J^1Y$ is defined by a form
\[
L=\cL(x^\la,y^i,y^i_\la)dx^1\w...\w dx^n, \qquad n=\dim X.
\]

Dynamics of the fermion-gravitation complex will be investigated in
Part II of the work. Its Lagrangian density on the configuration space
$J^1S$ is choosen in such a way that the associated
Euler-Lagrange operator $\cE_L$ reproduces the total Dirac operator
$\cD$ mentioned above. This Lagrangian density however is degenerate.

If a Lagrangian density is degenerate, the corresponding
Euler-Lagrange equations are underdetermined and need supplementary
gauge-type conditions. In gauge theory, they are the familiar gauge
conditions. In general case, the above-mentioned supplementary
conditions remain elusive. To describe constraint field systems, one
can use the covariant multimomentum Hamiltonian formalism where
canonical momenta correspond to derivatives of fields with respect
to all world coordinates, not only the time
\cite{car,gun,6sar,sard,lsar}. Given a fibred manifold $Y\to X$, the
corresponding multimomentum phase space is the Legendre manifold
\[
\Pi=\op\w^nT^*X\otimes TX\otimes VY
\]
endowed with the canonical coordinates $(x^\la,y^i,p^i_\la)$.

The feature of a tetrad gravitational field as a Higgs field consists
in the fact that, on the constraint space, its canonical momenta are
equal to zero, otherwise in the presence of fermion fields. Fermion
fields deform the constraint space in the gravitation sector, and
this deformation makes contribution into the energy-momentum
conservation law. In the framework of the multimomentum Hamiltonian
formalism, we have the fundamental identity whose restriction to a
constraint space can be treated as the energy-momentum conservation law
\cite{sard}. In Part II, percularity of
this conservation law in gravitation theory will be considered.

\section{Geometric Preliminary}

This Section aims to summarize the necessary prerequisites of jet
manifolds and connections \cite{man,sard,sau}.

All morphisms  throughout are differentiable maps of
class $C^\infty$.  Manifolds are real, Hausdorff,
finite-dimensional, second-countable and connected.

We use the standard symbols $\otimes$, $\wedge$ and $\rfloor$ for the
tensor, exterior and interior products respectively.

Given a manifold $Z$ with an atlas of local
coordinates $(z^\la)$,
the tangent bundle $TZ$ and the cotangent bundle
$T^*Z$ of  $Z$ are provided with the induced coordinates
$(z^\la, \dot z^\la)$ and $(z^\la,\dot z_\la)$ with respect
to the holonomic bases $\dr_\la$ and $dz^\la$. If
$f:Z\to Z'$ is a manifold map, by  $Tf:TZ\to TZ'$ is meant the
morphism tangent to $f$.

A fibred manifold $Y$ is defined to be a surjective submersion
\[
\pi :Y\to X
\]
where, unless otherwise stated, $X$ is an $n$-dimensional manifold.
A fibred manifold $Y$ is provided with an atlas of
fibred coordinates
\[
 (x^\la, y^i),\qquad x^\la \to {x'}^\la(x^\m), \qquad y^i \to
{y'}^i(x^\m,y^j).
\]
A locally trivial fibred manifold $Y$ is called a bundle, and its
fibred coordinates $y^i$ are coordinates of its standard fibre $V$.

 By a fibred morphism of a fibred manifold
$Y\to X$ to a fibred manifold $Y'\to X'$ is meant a fibre-to-fibre
manifold map $\Phi : Y\to Y'$ over a map $f: X\to X'$.
If $f=\Id_X$, one says briefly that $\Phi$ is a fibred
morphism $\op\to_X$ over $X$.

Given a fibred manifold $Y\to X$, every map
$f : X' \to X$ induces the pullback $f^*Y\to X'$ comprising the pairs
\[
\{(y,x')\in Y\times X' \mid \quad \pi(y) =
f(x')\}.
\]
In particular, the product
\[
Y=Y\op\times_X Y'
\]
of fibred manifolds over $X$ is defined.
For instance, given the tangent bundle $TX$ and the cotangent bundle
$T^*X$ of $X$, we have the products
\[
Y\op\times_X TX, \qquad  Y\op\times_X T^*X
\]
over $X$. For the sake of simplicity, we denote them  by the
symbols  $TX$ and $T^*X$ respectively.

The tangent bundle of a fibred manifold $Y$  contains
the vertical tangent subbundle
\[
VY = \Ker (T\pi)
\]
with the induced coordinates $(x^\la,y^i,\dot y^i).$
Given a fibred morphism $\Phi: Y\to Y'$, the tangent
morphism $T\Phi$ yields the corresponding vertical tangent morphism
$ V\Phi:VY\to VY'$.
With $VY$, we have the exact sequence
\begin{equation}
0\to VY\to TY\op\to_Y Y\op\times_X TX\to 0. \label{1.8a}
\end{equation}
Its different splittings
\[
Y\op\times_X TX\op\to_Y TY
\]
correspond to different connections on the fibred manifold $Y\to X$.

On fibred manifolds, we
consider  the following types of differential forms:

(i) exterior horizontal forms $ Y\to\op\w^r T^*X;$

(ii) tangent-valued horizontal forms $Y\to\op\w^r T^*X\op\otimes_Y TY$
and, in particular, soldering forms $Y\to T^*X\op\otimes_YVY$;

(iii) pullback-valued forms
\be
&&Y\to \op\w^r T^*Y\op\otimes_Y TX, \\
&&Y\to \op\w^r T^*Y\op\otimes_Y T^*X.
\ee

Horizontal $n$-forms are called horizontal densities.

Given a fibred manifold $Y\to X$, the first order jet manifold (or
simply the jet manifold) $J^1Y$ of $Y$
comprises the equivalence classes $j^1_xs$, $x\in X$, of sections $s$
of $Y$ so that different sections $s$ and $s'$ belong to the same jet
class $j^1_xs$ if and only if the tangent morphisms  $Ts$ and $Ts'$
consist with each other on the tangent
space $T_x$ to $X$. In other words, sections $s,s'\in j^1_xs$ are
identified by  their values $s^i(x)= {s'}^i(x)$ and values of their
partial derivatives $\dr_\m s^i(x) = \dr_\m {s'}^i(x)$ at $x$.
The jet manifold $J^1Y$ is provided with the adapted coordinates
\be
&&(x^\la,y^i,y_\la^i), \qquad y_\la^i(j^1_xs)=\dr_\la s^i(x),\\
&&{y'}^i_\la = (\frac{\dr{y'}^i}{\dr y^j}y_\m^j +
\frac{\dr{y'}^i}{\dr x^\m})\frac{\dr x^\m}{\dr{x'}^\la}.
\ee
It is both a fibred manifold $j^1_xs\to
x$ over  $X$ and an affine bundle  $j^1_xs\to s(x)$ over $Y$.
There exists the canonical  monomorphism
\be
&&\la:J^1Y\to T^*X \op\otimes_Y TY, \\
&& \la=dx^\la\otimes (\dr_\la + y^i_\la \dr_i),
\ee
over $Y$ whose image is an affine bundle modelled on the vector bundle
\begin{equation}
T^*X\op\otimes_Y VY. \label{L0}
\end{equation}
The monomorphism $\la$ is called the contact map. It enables us to handle
jets as tangent-valued forms.

Every fibred morphism $\Phi:Y\to Y'$ over a diffeomorphism of $X$
admits the jet prolongation
 \be
 &&J^1\Phi: J^1Y\to J^1Y', \\
&&{y'}^i_\la\circ J^1\Phi = (\frac{\dr\Phi^i}{\dr y^j}y_\m^j +
\frac{\dr\Phi^i}{\dr x^\m})\frac{\dr x^\m}{\dr{x'}^\la}.
\ee
In particular, every section $s$ of a fibred manifold $Y$ has the jet
prolongation
\[
(x^\la,y^i,y_\la^i)\circ J^1s= (x^\la,s^i(x),\dr_\la s^i(x))
\]
to the section $J^1s$ of the fibred jet manifold $J^1Y\to X$.

Note that the jet manifold machinery is naturally generalized to
complex bundles over a real base manifold.

Application of jet manifolds to differential geometry is based on the
horizontal splitting
\ben
&& J^1Y\op\times_Y TX \to  J^1Y\op\times_Y TY, \nonumber\\
&&\dot x^\la\dr_\la
+\dot y^i\dr_i =\dot x^\la(\dr_\la +y^i_\la\dr_i) + (\dot y^i-\dot x^\la
y^i_\la)\dr_i, \label{1.20}
\een
of the tangent bundle $TY$ over $J^1Y$.

A connection on a fibred manifold $Y\to X$ is defined
to be a global section
\be
&&\G :Y\to J^1Y,\\
&&(x^\la ,y^i,y^i_\la)\circ\G =(x^\la, y^i,\G^i_\la (y)),\\
&&{\G'}^i_\la = (\frac{\dr{y'}^i}{\dr y^j}\G_\m^j +
\frac{\dr{y'}^i}{\dr x^\m})\frac{\dr x^\m}{\dr{x'}^\la},
\ee
of the affine bundle $J^1Y\to Y$.
For instance, a linear connection on a vector bundle $Y$ reads
\[
\G^i_\la(y) = \G^i{}_{j\la}(x)y^j.
\]

By means of the contact map $\la$, any connection $\G$ on
$Y\to X$  can be represented by the
tangent-valued  1-form $\la\circ\G$ on $Y$ which we
denote by the same symbol
\begin{equation}
\G =dx^\la\otimes(\dr_\la +\G^i_\la (y)\dr_i). \label{37}
\end{equation}
Substituting (\ref{37}) into the canonical horizontal
splittings (\ref{1.20}), we obtain the familiar horizontal
splitting of the tangent bundle $TY$ with respect to a connection on $Y$.

Since $J^1Y\to Y$ is an affine bundle modelled on the vector bundle
(\ref{L0}), connections on a fibred manifold $Y$ constitute an affine
space modelled on the linear space of soldering forms on $Y$.
 Namely, if $\G$ is a connection and $\si$ is a soldering form on a fibred
manifold $Y$, then $\G+\si$ is a connection on $Y$. Conversely, if $\G$ and
$\G'$ are connections on a fibred manifold $Y$, then $\G-\G'$ is a soldering
form on $Y$.

Every connection $\G$ on a fibred manifold $Y$ yields the affine
bundle morphism
\ben
&&D_\G:J^1Y\op\to_Y T^*X\op\otimes_Y VY,\nonumber\\
&&D_\G =(y^i_\la -\G^i_\la)dx^\la\otimes\dr_i, \label{14}
\een
 which is called the covariant differential. The corresponding covariant
derivative  of a section $s$ of  $Y$ reads
\[
\nabla_\G s=D_\G\circ J^1s=(\dr_\la s^i-
(\G\circ s)^i_\la)dx^\la\otimes\dr_i.
\]
A section $s$ of a fibred manifold $Y$ is called an integral section for a
connection $\G$ on $Y$ if $\G\circ s=J^1s,$ that is, $\nabla_\G s=0$.

The general approach to connections as jet fields is suitable
to formulate the classical concept of  principal connections.

Let $P\to X$ be a principal bundle with a structure group $G$ which
is assumed to be a real finite-dimensional Lie group. By $r_g$, $g\in G$,
we denote the canonical action of $G$ on $P$ on the right.

In case of a principle bundle with a structure group $G$,
 the exact sequence (\ref{1.8a}) reduces to the exact sequence
\[
0\to V^GP\to T^GP\to TX\to 0
\]
 where
\[
 T^GP=TP/G,\qquad V^GP=VP/G
\]
are respectively the quotients of the tangent bundle $TP$ and the
vertical tangent bundle $VP$ of $P$ by the tangent and vertical tangent
morphisms to the canonical mappings $r_g$.

A principal connection $A$ on
a principal bundle $P\to X$ is defined to be a
 $G$-equivariant connection on $P$ such that
\be
&&A\circ r_g=J^1r_g\circ A, \qquad g\in G,\\
&& A=dx^\la\otimes(\dr_\la +A^m_\la(q)\tau_m), \qquad q\in P,\\
&& A^m_\la(qg)= A^m_\la(q) {\rm ad}g^{-1}(\tau_m),
\ee
where $\tau_m$ are the fundamental vector fields on $P$.
Given an atlas $\Psi^P=\{U_\xi, z_\xi\}$ of $P$
where $\{z_\xi\}$ is a family of local sections of $P$, we reproduce the
familiar local connection 1-forms
\begin{equation}
A_\xi=A^m_\la(x)dx^\la\otimes I_m \qquad A^m_\la(x)=A^m_\la(z_\xi (x)),
\label{1.32}
\end{equation}
where $I_m$ is a basis for the left Lie algebra of the group $G$.

There is the 1:1 correspondence between the principal connections on a
principal bundle $P\to X$  and the global sections of the quotient
\begin{equation}
C=J^1P/G\to X \label{68}
\end{equation}
of $J^1P$ by the jet prolongations of the canonical morphisms $r_g$.
We call $C$ the principal connection bundle. It is an affine
bundle modelled on the vector bundle
 \[
\ol C =T^*X \otimes V^GP.
\]

Given a bundle atlas $\Psi^P$ of $P$, the principal connection bundle $C$
is provided with  the fibred coordinates $(x^\mu,k^m_\mu)$ so that
\[
(k^m_\mu\circ A)(x)=A^m_\mu(x)
\]
 are coefficients of the local connection 1-form (\ref{1.32}).
The first order jet manifold $J^1C$ of  $C$ is endowed
 with the adapted coordinates $(x^\mu, k^m_\mu, k^m_{\mu\lambda})$.

Let $Y\to X$ be a bundle associated with a principal bundle $P\to X$.
The structure group $G$ of $P$ acts freely on the standard fibre $V$ of
$Y$ on the left. The total space of $Y$, by definition, is the quotient
\[
Y=(P\times V)/G
\]
of the product $P\times V$ by identification of elements $(qg\times gv)$
for all $g\in G$. The $P$ associated bundle $Y$
is provided with atlases $\Psi=\{U_\xi, \psi_\xi\}$
associated with atlases $\Psi^P=\{U_\xi, z_\xi\}$ of $P$ as follows:
\[
\psi^{-1}_\xi (x\times V)= [z_\xi (x)]_V (V), \qquad x\in U_\xi,
\]
where by $[q]_V$ is meant the restriction of the canonical map
$P\times V\to Y$ to $q\times V$.

Every principal connection $A$ on a principal bundle $P$ yields the
associated principal connection $\G$ on a $P$ associated bundle $Y$. With
respect to associated atlases $\Psi$ of $Y$ and $\Psi^P$ of $P$,
this connection reads
\[
\G=dx^\la\otimes (\dr_\la +A^m_\mu (x)I_m{}^i{}_jy^j\dr_i)
\]
where $A^m_\mu (x)$ are coefficients of the local connection 1-form
(\ref{1.32}) and $I_m$ are generators of the structure group $G$
on the standard fibre $V$ of the bundle $Y$.

\section{Composite Manifolds}

 A composite manifold
 is defined to be composition of surjective submersions
\begin{equation}
 \pi_{\Si X}\circ\pi_{Y\Si}:Y\to \Si\to X. \label{1.34}
\end{equation}
It is a fibred manifold $Y\to X$ provided with the
particular class of coordinate atlases
$( x^\la ,\si^m,y^i)$:
\[
{x'}^\lambda=f^\lambda(x^\mu), \qquad {\sigma'}^m=f^m(x^\mu,\sigma^n),
\qquad  {y'}^i=f^i(x^\mu,\sigma^n,y^j),
\]
 where $(x^\m,\si^m)$ are fibred  coordinates  of
$\Si\to X$. We further suppose that
$Y\to\Si$ is a bundle denoted by $Y_\Si$.

Application of composite manifolds to field theory is
based on the following assertions \cite{2sar,sard}.

(i) Let $Y$ be the composite manifold (\ref{1.34}).
Given a section $h$ of
$\Sigma\to X$ and a section $s_\Sigma$ of $Y\to\Sigma$, their
composition $s_\Sigma\circ h$ is a section of the composite manifold $Y$.
Conversely, every global section $s$ of the fibred manifold
$Y\to X$ is represented by some
composition $s_\Si\circ h$ where $h=\pi_{Y\Si}\circ s$ and $s_\Si$ is an
extension of the local section $h(X)\to s(X)$ of the bundle
$Y_\Si$ over the closed imbedded submanifold $h(X)\subset\Si$.

(ii) Given a global section $h$ of $\Sigma\to X$, the
restriction $Y_h=h^*Y_\Si$
of the bundle $Y\to\Sigma$ to $h(X)$ is a fibred imbedded submanifold
of $Y\to X$. Moreover, there is the 1:1 correspondence between the
sections of  $Y_h$ and the sections $s$ of
the composite manifold $Y$ such that $\pi_{Y\Sigma}\circ s =h$.

Therefore, one can say that sections $s_h$ of $Y_h\to X$
describe fields in the presence of a background field $h$, whereas sections
of the composite manifold $Y\to X$ describe all pairs $(s_h,h)$.
In accordance with the assertion (ii), there is the 1:1 correspondence
between them.
It is important when physical systems in the presence of
different background fields $h$ and $h'$ are nonequivalent  and can
not be represented by sections of the product $\Si\times Y$ where
$Y\to X$ is some standard bundle.

Let $Y_\Si$ be the composite  manifold (\ref{1.34}) and $J^1\Si$,
$J^1Y_\Si$ and $J^1Y$ the jet manifolds of
$\Si\to X$, $Y\to \Si$ and $Y\to X$ respectively. Given fibred
coordinates $(x^\la, \si^m, y^i)$ of $Y$,
the corresponding adapted coordinates of $J^1\Si$,
$J^1Y_\Si$ and $J^1Y$ are
\[
( x^\la ,\si^m, \si^m_\la),\qquad
( x^\la ,\si^m, y^i, \wt y^i_\la, y^i_m),\qquad
( x^\la ,\si^m, y^i, \si^m_\la ,y^i_\la).
\]

There exists  the  canonical surjection
\ben
&&\rho : J^1\Si\op\times_\Si J^1Y_\Sigma\op\to_Y J^1Y,\nonumber \\
&&\rho(j^1_xh,j^1_{h(x)}s_\Sigma)=j^1_x({s_\Sigma}\circ h), \label{1.38}\\
&&y^i_\lambda\circ\rho=y^i_m{\sigma}^m_{\lambda} +\wt y^i_{\lambda},
\nonumber
\een
where $s_\Si$ and $h$ are sections of $Y\to\Si$ and $\Si\to X$
respectively \cite{sau}.

In particular, let
\ben
&&\Gamma=dx^\lambda\otimes(\dr_\lambda + \Gamma^m_\lambda\dr_m),
\nonumber\\
&& A_\Sigma=dx^\lambda\otimes(\dr_\lambda+\wt A^i_\lambda\dr_i)
+ d\sigma^m\otimes(\dr_m+A^i_m\dr_i) \label{L2}
\een
be connections on fibred manifolds $\Sigma\to X$ and $Y\to\Si$
respectively. Building on the canonical morphism (\ref{1.38}),
one can construct the composite connection
\begin{equation}
A=dx^\lambda\otimes[\dr_\lambda+\Gamma^m_\lambda
\dr_m + (A^i_m\Gamma^m_\lambda + \wt A^i_\lambda)\dr_i] \label{1.39}
\end{equation}
on $Y\to X$ \cite{2sar,sard}.
Composite connections (\ref{1.39}) are by no means the unique type of
connections on a composite manifold. We consider them since,
if a connection $\Gamma$ on $\Si\to X$ has an integral section $h$,
the composite connection (\ref{1.39}) on $Y$ is reducible to the connection
\begin{equation}
A_h=dx^\lambda\otimes[\dr_\lambda+
(A^i_m\dr_\la h^m + \wt A^i_\lambda)\dr_i] \label{1.42}
\end{equation}
on the fibred submanifold $Y_h$ of $Y\to X$.

\section{Spontaneous Symmetry Breaking}

In classical field theory, spontaneous symmetry breaking is modelled
by classical Higgs fields. In geometric
terms, the necessary condition of spontaneous symmetry breaking consists
in reduction of a structure group $G$ of a principal bundle $P$ to
its closed subgroup $K$ of exact symmetries. Classical
Higgs fields are represented by global sections of the bundle
$P/K\to X$ or isomorphic homogeneous bundles.

Let $\pi_P:P\to X$ be a principal bundle with a structure
Lie group $G$ and $K$ its closed subgroup. Then $\Si=P/K\to X$
is the associated bundle
\[
P/K=(P\times G/K)/G
\]
with the standard
fibre $G/K$  on which the structure group $G$ acts on the left.
We have the composite manifold
\begin{equation}
\pi_{\Si X}\circ\pi_{P\Si}:P\to P/K\to X \label{7.16}
\end{equation}
 where $P_\Si=P\to P/K$ is a principal bundle with the structure group
$K$. Note that (\ref{7.16}) fails to be a principal bundle.

Let the structure group $G$ be reducible to its closed subgroup $K$. By the
well-known theorem, there is the 1:1 correspondence
\[
\pi_{P\Si}(P_h)=(h\circ\pi_P)(P_h)
\]
 between global sections $h$ of the
bundle $P/K\to X$ and the $K$ reduced subbundles $P_h$ of $P$ which
are restrictions of the principal bundle $P_\Si$ over $h(X)$.
Every principal connection $A_h$ on a reduced subbundle
$P_h$ is lifted to a principal connection on $P$. Conversely, a
principal connection $A$ on $P$ is reducible to a principal connection
on $P_h$ if and only if the global section $h$ of
the bundle $P/K\to X$ is an integral section of the connection $A$.

One says that sections $s_h$ of a vector bundle $Y_h\to X$ with a
standard fibre $V$ describe matter fields in the presence of a Higgs field
$h$ if $Y_h$ is associated with the reduced subbundle $P_h$
of the principal bundle $P$, that is,
\begin{equation}
Y_h=(P_h\times V)/K. \label{7.17}
\end{equation}
Matter fields $s_h$ in the presence of different Higgs fields $h$ and
$h'$ are described by sections of the matter bundles $Y_h$
and $Y_{h'}$ associated with different reduced subbundles
$P_h$ and $P_{h'}$ of $P$. If the standard fibre $V$ admits only
represenation of the exact symmetry subgroup $K\subset G$,
there is no canonical isomorphism between $Y_h$ and $Y_{h'}$.
In this case, a connection on $Y_h$ is assumed
to be associated with a principal
connection on the reduced subbundle $P_h$.
A principal connection $A_h$ on $P_h$ is extended to a principal
connection  on  $P$ which however fails to be
reducible to a connection on another reduced subbundle $P_{h'\neq h}$.
 It follows that matter
fields and gauge potentials possessing only exact symmetries must be
regarded only in pairs with a certain Higgs field.

To describe this spontaneous symmetry breaking, composite
manifolds have been suggested \cite{2sar}.

Given the composite manifold (\ref{7.16}), the
canonical morphism (\ref{1.38}) results in the surjection
\begin{equation}
J^1P_\Si/K\op\times_\Si J^1\Si \to J^1 P/K \label{7.18}
\end{equation}
over $J^1\Si$. In particular, let
$A_\Sigma$ be a principal connection on $P_\Si$
 and $\Gamma$ be a connection on $\Si$. The corresponding
composite connection (\ref{1.39}) on the composite manifold (\ref{7.16})
is equivariant under the canonical
action of $K$ on $P$. If a connection $\Gamma$ has an integral
global section $h$ of $P/K\to X$, the composite connection (\ref{1.39})
is reducible to the connection (\ref{1.42}) on $P_h$ which is a
principal connection on $P_h$.

Let us consider the composite manifold
\begin{equation}
Y=(P\times V)/K\to P/K\to X \label{7.19}
\end{equation}
 where the bundle
\[
Y_\Sigma = (P\times V)/K\to P/K
\]
is associated with the $K$ principal bundle
$P\to P/K$. Given a reduced subbundle $P_h$ of $P$, the associated bundle
\[
Y_h=(P_h\times V)/K
\]
is isomorphic to the restriction of $Y_\Sigma$ over $h(X)$.
The composite manifold (\ref{7.19}) can be provided with
the composite connection (\ref{1.39}) where the connection $A_\Si$ on the
bundle $Y\to P/K$ is a principal connection on the $K$ principal bundle
$P_\Si$ and the connection $\G$ on $\Si$ is
a principal connection on some reduced subbundle $P_h$ of $P$
\cite{2sar,sard}. In this case, the
composite connection $A$ is reducible to the connection (\ref{1.42})
on the bundle $Y_h$ (\ref{7.17}) as a subbundle of the composite
manifold $Y$. This connection appears to be some
principal connection $A_h$ on $P$.

Thus, they are sections of the composite manifold (\ref{7.19}) which
describe the above-mentioned pairs $(s_h,h)$ of matter fields
possessing only exact symmetries and Higgs fields.

Accordingly, the pairs $(A_h,h)$ of
exact symmetry gauge potentials $A_h$ and Higgs fields $h$
can be represented by sections of
the composite manifold
\begin{equation}
C_K= J^1P/K\to P/K\to X \label{N53}
\end{equation}
 which  are projected onto
jet prolongations $J^1h$ of global sections $h$ of the bundle $\Si$.

 Note that the manifold $(P\times V)/K$ possesses the structure of the
bundle associated with the principal bundle $P$. Its standard fibre
is $(G\times V)/K$ on which the structure group $G$ of $P$ (and its
subgroup $K$) acts by the law
\[
G\ni g: (G\times V)/K\to (gG\times V)/K.
\]
However it differs from the action of the structure group $K$ of $P_\Si$
on it by the law
\[
K\ni g: (G\times V)/K\to (Gg^{-1}\times V)/K.
\]

\section{Dirac Fermion Fields}

Spontaneous symmetry breaking in gravitation theory where Dirac fermion
fields possess only Lorentz symmetries belongs to the type mentioned
in previous Section.

By $X^4$ is further meant an oriented world manifold which satisfies
the well-known global topological conditions in order that gravitational
fields, space-time structure and spinor structure can exist. To
summarize these conditions, we assume that $X^4$ is not compact and
the linear frame bundle $LX$ is trivial \cite{3sar}.

We describe Dirac fermion fields as follows. Given
a Minkowski space $M$ with the Minkowski metric $\eta$, let
\[
A_M=\op\oplus_nM^n,\qquad M^0={\bf R},\qquad
M^{n>0}=\op\otimes^nM,
\]
be the tensor algebra modelled on $M$. The complexified quotient of this
algebra by the two-sided ideal generated by elements
\[
e\otimes e'+e'\otimes e-2\eta(e,e')\in A_M,\qquad e\in M,
\]
constitutes the complex Clifford algebra ${\bf C}_{1,3}$.
A spinor space $V$ is defined to be a
linear space of some minimal left ideal of ${\bf C}_{1,3}$  on
which this algebra acts on the left. We then have the representation
\begin{equation}
\gamma: M\otimes V \to V, \label{2.1}
\end{equation}
of elements of the Minkowski space $M\subset{\bf C}_{1,3}$ by
$\gamma$-matrices on $V$:
\[
\gamma (e^a\otimes y^Av_A) = \gamma^{aA}{}_By^Bv_A,
\]
where
$\{e^0...e^3\}$ is a fixed basis for $M$, $v_A$ is a basis for $V$,
and $\gamma^a$ are Dirac's matrices of a fixed form \cite{bug}.

Let us consider the transformations preserving the representation
(\ref{2.1}).
These are pairs $(l,l_s)$ of Lorentz transformations $l$ of  the Minkowski
space $M$ and invertible elements $l_s$ of ${\bf C}_{1,3}$ such that
\be
&&lM = l_s Ml^{-1}_s,\\
&&\gamma (lM\otimes l_sV) = l_s\gamma (M\otimes V).
\ee
Elements $l_s$  form the Clifford group whose action on $M$
however is not effective. We here  restrict ourselves to its spinor
subgroup
\[
L_s =SL(2,{\bf C}), \qquad L=L_s/{\bf Z}_2,
\]
whose generators act on $V$ by the representation
\[
I_{ab}=\frac{1}{4}[\gamma_a,\gamma_b].
\]

Let us consider a bundle of complex Clifford algebras ${\bf C}_{3,1}$
over $X^4$. Its subbundles are both a spinor bundle $S_M\to X^4$ and the
bundle $Y_M\to X^4$ of Minkowski spaces of generating elements of
${\bf C}_{3,1}$.
To describe Dirac fermion fields on a world manifold, one must
require $Y_M$ be isomorphic to the cotangent bundle $T^*X$
of a world manifold $X^4$. It
takes place only if the structure group of the principal linear frame
bundle $LX$ is reducible to the Lorentz group $L$ and $LX$
contains a reduced $L$ subbundle $L^hX$ such that
\[
Y_M=(L^hX\times M)/L=T^*X.
\]
In this case, the spinor bundle $S_M$ is associated with the $L_s$ lift
$P_h$ of $L^hX$:
\[
r:P_h\to L^hX=P_h/{\bf Z}_2,
\]
\[
S_M=(P_h\times V)/L_s=S_h.
\]

The geometric equivalence principle thereby is the necessary condition
in order that Dirac fermion fields can be defined on a world manifold.

There is the above-mentioned 1:1 correspondence between the reduced
subbubdles $L^hX$ of $LX$ and
the tetrad gravitational fields $h$ identified with global sections
of the bundle $\Si$ (\ref{5.15}).

Given a tetrad field $h$, let $\Psi^h=\{z^h_\xi\}$ be an atlas of
$LX$ which is extention of an atlas of $L^hX$, that is, the local
sections $z_\xi^h$ take their values into $L^hX$. With respect to
$\Psi^h$, the pseudo-Riemannian metric $g$ associated with a
gravitational field $h$ comes to the Minkowski metric which exemplifies
a Lorentz invariant defined in accordance with the geometric equivalence
principle.

With respect to an atlas $\Psi^h$ and a
holonomic atlas $\Psi^T=\{\psi_\xi^T\}$ of $LX$, a tetrad field $h$
can be represented by a family of $GL_4$-valued tetrad functions
\[
h_\xi=\psi^T_\xi\circ z^h_\xi. \qquad h=\pi_{P\Si}\circ z^h_\xi.
\]
They are $GL_4$-valued functions of atlas transformations
\begin{equation}
dx^\la= h^\la_a(x)h^a \label{L6}
\end{equation}
between the bases $dx^\la$ and $h^a$ for the cotangent spaces to
$X^4$ which are associated with atlases $\Psi^T$ and $\Psi^h$
respectively.

Given a tetrad field $h$, one can define the representation
\begin{equation}
\gamma_h: T^*X\otimes S_h= (P_h\times (M\otimes V))/L_s\to (P_h\times
\gamma(M\otimes V))/L_s = S_h \label{L4}
\end{equation}
of cotangent vectors to a world manifold $X^4$ by Dirac's $\g$-matrices
on elements of the spinor bundle $S_h$. With respect to an atlas
$\{z_\xi\}$ of $P_h$ and the associated atlas $\{z^h_\xi=r\circ
z_\xi\}$ of $LX$, the morphism (\ref{L4}) reads
\[
\g_h(h^a\otimes y^Av_A(x))=\g^{aA}{}_By^Bv_A(x)
\]
where $\{h^a\}$ and $\{v_A(x)\}$ are the associated bases for
fibres $T^*_xX$ of $T^*X$ and $V_x$ of $S_h$ respectively. As a
shorthand, we can write
\be
&&\wh h^a =\g_h(h^a)= \g^a.\\
&& \wh dx^\la=\g_h(dx^\la)=h^\la_a(x)\g^a.
\ee

We shall say that, given the representation (\ref{L4}), sections of
the spinor bundle $S_h$ describe fermion fields in the presence of
the tetrad gravitational field $h$.

Let $A_h$ be a principal connection on $S_h$ and
\be
&&D: J^1S_h\op\to_{S_h}T^*X\op\otimes_{S_h}VS_h,\\
&&D=(y^A_\la-A^{ab}{}_\la (x)I_{ab}{}^A{}_By^B)dx^\la\otimes\dr_A,
\ee
the corresponding covariant differential (\ref{14}). Given the
representation (\ref{L4}), one can construct the Dirac operator
\begin{equation}
 \cD_h=\g_h\circ D: J^1S_h\to T^*X\op\otimes_{S_h}VS_h\to VS_h, \label{I13}
\end{equation}
\[
\dot y^A\circ\cD_h=h^\la_a(x)\g^{aA}{}_B(y^B_\la-A^{ab}{}_\la
I_{ab}{}^A{}_By^B).
\]
We use the fact that the vertical tangent bundle $VS_h$ admits the
canonical splitting
\[
VS_h=S_h\times S_h,
\]
and $\g_h$ in the expression (\ref{I13}) is the pullback
\be
&& \g_h: T^*X\op\otimes_{S_h}VS_h\op\to_{S_h}VS_h,\\
&& \g_h(h^a\otimes\dot y^A\dr_A)=\g^{aA}{}_B\dot y^B\dr_B,
\ee
over $S_h$ of the bundle morphism (\ref{L4}). On sections $\phi_h$ of
$S_h$, we have the familiar expression
\[
\cD_h\circ\phi_h=\wh dx^\la\nabla_\la\phi_h = h^\la_a(x)\wh h^a
\nabla_\la\phi_h =h^\la_a(x)\g^a\nabla_\la\phi_h
\]
for the Dirac operator in the presence of a tetrad gravitational field $h$.

 For different tetrad fields $h$ and $h'$, Dirac fermion fields are
described by sections of spinor bundles $S_h$ and $S_{h'}$ associated
with $L_s$ lifts $P_h$ and $P_{h'}$ of different reduced $L$ subbundles
of $LX$. Threfore, the representations $\gamma_h$ and $\gamma_{h'}$
(\ref{L4}) are not equivalent \cite{tsar,3sar}.

For two arbitrary elements $q\in P_h$ and $q'\in P_{h'}$ over the same
point $x\in X$, there is an element $g\in GL_4$ so that
\[
rq'=(r_g\circ r)q.
\]
Let $T_x^*$ be the cotangent space to $X^4$ at $x\in X^4$. Since
\[
T^*_x=[rq]_MM=([rq']_M\circ g^{-1})M,
\]
we can write
\be
&& \g_h: T^*_x\otimes V_x=[rq]_MM\otimes[q]_VV\to([q]_V\circ\g)
(M\otimes V),\\
&& \g_{h'}: T^*_x\otimes {V'}_x=([rq]_M\circ g)M\otimes[q']_VV\to([q']_V
\circ\g)(gM\otimes V).
\ee
If $g\in GL_4\setminus L$, there is no isomorphism $l_V$ of the spinor
space $V$ such that
\[
\g(gM\otimes l_VV)= l_V\g(M\otimes V).
\]

It follows that a Dirac fermion field must be regarded only in a pair with
a certain tetrad gravitational field. There is the 1:1 correspondence
between these pairs and sections of the composite bundle (\ref{L1}).

\section{Composite Spinor Bundles}

In gravitation theory, we have the composite manifold
\begin{equation}
\pi_{\Si X}\circ\pi_{P\Si}:LX\to\Si\to X^4 \label{L3}
\end{equation}
where $\Si$ is the bundle $LX/L\to X^4$ (\ref{5.15}) and $LX_\Si=LX\to\Si$
is the L principal bundle.

Since the fibration
\[
GL_4\to GL_4/L
\]
is trivial, the bundle $LX_\Si$ is isomorphic to the pullback of some
$L$ principal bundle over $X^4$ by $\pi_{\Si X}$. There exists the
principal $L_s$ lift $P_\Si$ of $LX_\Si$ such that
\[
P_\Si/L_s=\Si, \qquad LX_\Si=rP\Si=P\Si/{\bf Z}_2.
\]
In particular, there is imbedding of $P_h$ onto the restriction of $P_\Si$
onto $h(X^4)$.

Let us consider the composite spinor bundle (\ref{L1})
\begin{equation}
S=\pi_{\Si X}\circ\pi_{S\Si}: (P_\Si\times V)/L_s\to\Si\to X^4 \label{L5}
\end{equation}
where $S_\Si=S\to\Si$ is
associated with the $L_s$ principal bundle $P_\Si$. It is readily observed
that, given a global section $h$ of $\Si\to X^4$, the restriction $S\to
\Si$ to $h(X^4)$ is the spinor bundle $S_h$ whose sections describe Dirac
fermion fields in the presence of the tetrad field $h$.

Let us provide the principal bundle $LX$ with a holonomic atlas
$\{\psi^T_\xi, U_\xi\}$ and the principal bundles $P_\Si$ and $LX_\Si$
with associated atlases $\{z^s_\e, U_\e\}$ and $\{z_\e=r\circ z^s_\e\}$.
With respect to these atlases, the composite spinor bundle is endowed
with the fibred coordinates $(x^\la,\si_a^\mu, y^A)$ where $(x^\la,
\si_a^\mu)$ are fibred coordinates of the bundle $\Si\to X$ such that
$\si^\mu_a$ are the matrix components of the group element
\[
GL_4\ni (\psi^T_\xi\circ z_\e)(\si): {\bf R}^4\to {\bf R}^4,
\qquad \si\in U_\e,\qquad \pi_{\Si X}(\si)\in U_\xi.
\]
Given a section $h$ of $\Si\to X^4$, we have
\be
&&z^h_\xi (x)= (z_\e\circ h)(x), \qquad h(x)\in U_\e,
\qquad x\in U_\xi,\\
&& (\si^\la_a\circ h)(x)= h^\la_a(x),
\ee
where $h^\la_a(x)$ are tetrad functions (\ref{L6}).

The jet manifolds $J^1\Si$, $J^1S_\Si$ and $J^1S$ are provided with the
adapted coordinates
\be
&&(x^\la,\si^\mu_a, \si^\mu_{a\la}),\\
&&(x^\la,\si^\mu_a, y^A,\wt y^A_\la, y^A{}^a_\mu),\\
&&(x^\la,\si^\mu_a, y^A,\si^\mu_{a\la}, y^A_\la).
\ee
Note that, for each section $h$ of $\Si$, the fibred jet manifold
$J^1S_h\to X^4$ is a fibred submanifold of $J^1S$ given by the coordinate
relations
\[
\si^\mu_a=h^\mu_a(x), \qquad \si^\mu_{a\la}=\dr_\la h^\mu_a(x).
\]

Let us consider the bundle of Minkowski spaces
\[
(LX\times M)/L\to\Si
\]
associated with the $L$ principal bundle $LX_\Si$. It is isomorphic to
the pullback $\Si\times T^*X$ which we denote by the same symbol
$T^*X$. Building on the morphism (\ref{2.1}), one can define the bundle
morphism
\begin{equation}
\g_\Si: T^*X\op\otimes_\Si S_\Si= (P_\Si\times (M\otimes V))/L_s
\to (P_\Si\times\g(M\otimes V))/L_s=S_\Si \label{L7}
\end{equation}
over $\Si$. In the coordinate form, we have
\[
\wh dx^\la=\g_\Si (dx^\la) =\si^\la_a\g^a
\]
where $dx^\la$ is the basis for the fibre of $T^*X$ over $\si\in\Si$.
Being restricted to the submanifold $h(X^4)\subset \Si$ for a section
$h$ of $\Si$, the morphism (\ref{L7}) comes to the morphism $\g_h$
(\ref{L4}). Because of the canonical vertical splitting
\[
VS_\Si =S_\Si\op\times_\Si S_\Si,
\]
the morphism (\ref{L7}) yields the corresponding morphism
\begin{equation}
\g_\Si: T^*X\op\otimes_SVS_\Si\to VS_\Si. \label{L8}
\end{equation}

We use this morphism in order to construct the total Dirac
operator on sections of the composite spinor bundle (\ref{L5}). We are
based on the following fact.

Let $Y\to\Si\to X$ be a composite manifold (\ref{1.34}). Every connection
(\ref{L2}) on the bundle $Y\to\Sigma$ defines the splitting
\[
VY=VY_\Sigma\op\oplus_Y (Y\op\times_\Sigma V\Sigma),
\]
\[
\dot y^i\dr_i+\dot\si^m\dr_m = (\dot y^i-A^i_m\dot\si^m)\dr_i
+\dot\si^m(\dr_m+A^i_m\dr_i),
\]
of the vertical tangent bundle $VY$ to $Y\to X$. As a consequense,
we have  the following morphism over $Y$:
\ben
&&\wt D: J^1Y\to T^*X\op\otimes_Y VY_\Si, \nonumber\\
 &&\wt D= dx^\la\otimes(y^i_\la-\wt A^i_\la -A^i_m\si^m_\la)\dr_i.
\label{L9}
\een

Let
\begin{equation}
\wt A=dx^\la\otimes (\dr_\la +\wt A^B_\la\dr_B) + d\si^\mu_a\otimes
(\dr^a_\mu+A^B{}^a_\mu\dr_B) \label{L10}
\end{equation}
be a connection on the bundle $S_\Si$. Then, the composition of
morphisms (\ref{L8}) and (\ref{L9}) results in the following morphism
over $S$:
\begin{equation}
\cD=\g_\Si\circ\wt D:J^1S\to T^*X\op\otimes_SVS_\Si\to VS_\Si,\label{L11}
\end{equation}
\[
\dot y^A\circ\cD=\si^\la_a\g^{aA}{}_B(y^B_\la-\wt A^B_\la -
A^B{}^a_\mu\si^\mu_{a\la}).
\]
One can treat this morphism as the total Dirac operator since, for each
tetrad field $h$, the restriction of $\cD$ to $J^1S_h\subset J^1S$ comes
to the Dirac operator $\cD_h$ (\ref{I13}) in the presence of a
principal connection
\[
A_h=dx^\la\otimes[\dr_\la+(\wt A^B_\la+A^B{}^a_\mu\dr_\la h^\mu_a)\dr_B].
\]
The Dirac operator (\ref{L11}) thus characterizes the fermion-gravitation
complex only in the presence of background gauge gravitational potentials
$A_h$.

In the gauge gravitation theory, classical gravity is described by pairs
$(h,A_h)$ of tetrad gravitational fields $h$ and gauge gravitational
potentials $A_h$ identified with principal connections on the reduced
$L$ subbundles $L^hX$ of $LX$. Every connection on $L^hX$ is extended
to a Lorentz connection on $LX$ which however fails to be reducible
to a principal connection on another reduced subbundle $L^{h'}X$ if
$h\neq h'$. It follows that gauge gravitational potentials also must be
regarded in pairs with a certain tetrad gravitational field $h$.
Following Section 4, one can describe these pairs $(h,A_h)$ by sections
of the bundle (\ref{N53}) where $P=LX$ and $K=L$. The corresponding
configuration space is the jet manifold $J^1C_L$.

The bundle $C_L=J^1(LX)$ is endowed with the local fibred coordinates
\[
(x^\mu,\sigma^\mu_a,k^{ab}{}_\lambda=
-k^{ba}{}_\lambda, \sigma^\mu_{a\la})
\]
where $(x^\mu,\sigma^\mu_a, \sigma^\mu_{a\la})$ are coordinates
of the jet bundle $J^1\Si$.
Given a section $s$ of $C_L$, we have familiar tetrad functions and Lorentz
gauge potentials
\[
(\sigma^\mu_a\circ s)(x)=h^\mu_a(x), \qquad (k^{ab}{}_\lambda \circ
s)(x)=A^{ab}{}_\lambda(x)
\]
respectively. The jet manifold $J^1C_L$ of $C_L$ is provided with
the adapted coordinates
\[
(x^\mu,\sigma^\mu_a,k^{ab}{}_\lambda=
-k^{ba}{}_\lambda, \sigma^\mu_{a\la}=\sigma^\mu_{a(\la)},
k^{ab}{}_{\mu\lambda},\sigma^\mu_{a\la\nu}).
\]

The total configuration space of the fermion-gravitation complex is
the product
\[
J^1C_L\op\times_{J^1\Si}J^1S.
\]
Part II of the work will be devoted to dynamics of the fermion-gravitation
complex on this configuration space and the associated phase space.

\end{document}